\documentclass{ws-ijmpc}

\begin{document}

\markboth{  J. Seke, A.V. Soldatov, M. Polak, G. Adam,}
{Computationally efficient analytic representation  ... }

\title{\phantom{on}}

\title{Computationally efficient analytic representation of relativistic transition matrix elements in the Lamb shift calculations for hydrogenic atoms}

\author{J. SEKE}

\address{Institut f\"ur Theoretische Physik, Technische Universit\"at
Wien\\ Wiedner Hauptstrasse 8-10/136, A-1040 Wien, Austria\\
jseke@tph.tuwien.ac.at}

\author{A.V. SOLDATOV }

\address{Department of Mechanics, V.A. Steklov Mathematical Institute\\ of
the Russian Academy of Sciences,\\ 8, Gubkina str., Moscow,
119991, Russia
\\ soldatov@mi.ras.ru}

\author{M. POLAK}

\address{Institut f\"ur Theoretische Physik, Technische Universit\"at
Wien\\ Wiedner Hauptstrasse 8-10/136, A-1040 Wien, Austria\\
mpolak@tph.tuwien.ac.at}

\author{G. ADAM}

\address{Institut f\"ur Theoretische Physik, Technische Universit\"at
Wien\\ Wiedner Hauptstrasse 8-10/136, A-1040 Wien, Austria\\
gadam@tph.tuwien.ac.at}

\maketitle

\begin{abstract}
By using the plane-wave expansion for the electromagnetic-field
vector potential, transition matrix elements between the
relativistic bound and unbound states of hydrogenic atoms  were
expressed explicitly in terms of finite series made of
hypergeometric functions of the type $\phantom{ }_2F_1$. This
general formulae representation for the above mentioned matrix
elements in terms of hypergeometric functions proved very
convenient for direct numerical calculation of their contributions
to the Lamb shift in hydrogenic atoms. All integrations over the
angular variables of the wave vector ${\bf k}$ can be carried out
analytically with the exception of only one principal-value
integration over the absolute value $|{\bf k}|$ is left to be
carried out numerically. Conciseness and reliance on functions
already built-in to the standard computational packages like {\em
Mathematica} renders this approach  highly favorable for
programming of computationally efficient algorithms.
\end{abstract}



\newcommand{\vp}{\vec p\,}
\newcommand{\dsum}{\displaystyle\sum}
\newcommand{\dint}{\displaystyle\int}
\newcommand{\hb}{\hbar}

\newcommand{\noi}{\noindent}
\newcommand{\fr}{\frac}

\newcommand{\vx}{{\bf x}}
\newcommand{\tm}{\tilde m}
\newcommand{\om}{\omega}

\newcommand{\Om}{\Omega}

\newcommand{\lam}{\lambda}

\newcommand{\Gam}{\Gamma}

\newcommand{\gam}{\gamma}

\newcommand{\ran}{\rangle}

\newcommand{\lan}{\langle}

\newcommand{\da}{\dagger}

\newcommand{\del}{\delta}

\newcommand{\inft}{\infty}

\newcommand{\Del}{\Delta}

\newcommand{\bg}{\begin{equation}}

\newcommand{\en}{\end{equation}}

\newcommand{\lb}{\label}

\newcommand{\pz}{P^{(0)}}

\newcommand{\po}{P^{(1)}}

\newcommand{\pt}{P^{(2)}}

\newcommand{\sig}{\sigma}

\newcommand{\Sig}{\Sigma}

\newcommand{\ba}{\begin{eqnarray}}

\newcommand{\bac}{\begin{array}{c}}

\newcommand{\eac}{\end{array}}

\newcommand{\ea}{\end{eqnarray}}

\newcommand{\ds}{\displaystyle\sum}

\newcommand{\ce}{{\cal E}}

\newcommand{\di}{\displaystyle\int}

\newcommand{\vr}{{\bf r}}

\newcommand{\vk}{{\bf k}}

\newcommand{\al}{\alpha}

\newcommand{\nn}{\nonumber}

\newcommand{\bt}{\begin{tabular}}

\newcommand{\et}{\end{tabular}}

\newcommand{\ka}{\kappa}

\newcommand{\ep}{\epsilon}

\newcommand{\til}{\tilde}

\newcommand{\veps}{\varepsilon}

\renewcommand{\vec}{\mathbf}



\let\a=\alpha

\let\b=\beta

\let\g=\gamma

\let\d=\delta

\let\e=\epsilon

\let\ve=\varepsilon

\let\z=\zeta

\let\h=\eta

\let\th=\theta

\let\dh=\vartheta

\let\k=\kappa

\let\l=\lambda

\let\m=\mu

\let\n=\nu

\let\x=\xi

\let\p=\pi

\let\r=\rho

\let\s=\sigma

\let\t=\tau

\let\o=\omega

\let\c=\chi

\let\ps=\psi

\let\ph=\varphi

\let\Ph=\phi

\let\PH=\Phi

\let\Ps=\Psi

\let\O=\Omega

\let\S=\Sigma

\let\P=\Pi

\let\Th=\Theta

\let\L=\Lambda

\let\G=\Gamma

\let\D=\Delta

\let\X=\Xi

\let\u=\underline

 \def\<{\langle } \def\>{\rangle }

 \def\ri {{\ri}}



\section{Introduction}

 The primary  aim of the ongoing research is to develop a numerical calculation
technique allowing for practical calculation of renormalized expression for the Lamb shift in hydrogenic atoms obtained
 by  consistent  renormalization in QED developed by Seke
\cite{Sec1,Sec2}. Employing this method, the usage of the plane
vector expansion for the vector potential is a preferable choice.
In the context of this ultimate goal, plain-style Lamb-shift
calculation undertaken in the present study, besides being of
value in itself, also provides a good testing ground for the
techniques of analytical and numerical evaluation of transition
matrix elements between bound-bound and bound-unbound eigenstates
of hydrogenic atoms elaborated so far just for the case of the
plane-wave expansion for the electromagnetic-field vector
potential\cite{hypersecp,hyperfirstp,evtrmatel,fourtr}.

In the second quantization, the relativistic Dirac Hamiltonian for the
interaction between the atom and the radiation field reads as

\begin{eqnarray}
H_{\rm I}&=&\int d^3rH_{\rm I}(\vec{r})=
-e\int d^3r j^{\m}(\vec{r})A_{\m}(\vec{r})\\
j^{\m}(\vec{r})&=&\frac{1}{2}[\overline{\psi}(\vec{r}),\g^{\m}
\psi(\vec{r})]
\label{DH}
\end{eqnarray}

\noi with the vector potential $A_{\m}(\vec{r})$ in the covariant
quantization (Feynman gauge):

\begin{eqnarray}
 A_{\m}(\vec{r}) =
 \frac{1}{\sqrt{2(2\pi)^3}} \sum_{\l=0}^{3} \int d^3k (k)^{-1/2}
[ a^{\l,-}_{\vec{k}}\e_{\m}^{\l}(\vec k)e^{-i\vec{k}\vec{r}} +
a^{\l,+}_{\vec{k}}\e_{\m}^{\l}(\vec k)e^{i\vec{k}\vec{r}}].
\label{vp}
\end{eqnarray}

\noi Here, $\g^{\mu}$ are the gamma matrices,
$a^{\l,\pm}_{\vec{k}},\l=0,1,2,3$ are the photon creation and
annihilation operators for the mode $\vec{k},\l$, and
$\psi^{\eta}(\vec r)$ and $\overline{\psi}^{\,\eta}(\vec
x)=[\psi^{\eta}(\vec r)]^+\g_0$ are the fermion field and its
Dirac adjoint:

\begin{eqnarray}
\psi(\vec{x})= \sum_{s}c_{s}^- u_{s}(\vec{r})+\sum_{s'}
d_{s'}^+ v_{s'}(\vec{r}).
\end{eqnarray}

Here, $s$  stands  for both the set of quantum numbers of the
discrete and that of the continuous spectrum, while $s'$ stands for the set of quantum numbers of the continuous spectrum only, $c^{\pm}_{s}$ and
$d_{s'}^{\pm}$ are the electron and positron creation and
annihilation operators with the corresponding Dirac eigenfunctions
$u_{s}$ and $v_{s'}$. This yields the following transition
matrix elements
\ba M_{s_1,s_2}({\bf k},\lambda)&=&M(s_1,s_2,{\bf
k}, \lambda)= \overline{\langle s_1; vac|}H_I|s_2; {\bf k},\l
\rangle= \nn \\
&=&-e\di d^3 r \overline{\langle s_1|}j^{\m}({\bf r})|s_2\rangle
\langle vac| A_{\m}({\bf r})|{\bf k},\l\rangle.
\label{transmatrel} \ea

\noi Here,  for the discrete spectrum, the  discrete-spectrum eigenstates
$|s_1\rangle$ and $|s_2\rangle$ of the hydrogen-like atom are denoted by
the quantum numbers $s_i=\{n'_i,\kappa_i,m_i\}; \, i=1,2$, where
$n'_i$ is the radial quantum number and $m_i$ is the magnetic
quantum number. The eigenstates belonging to the continuous spectrum are denoted by the quantum numbers $s_i=\{p_i,\kappa_i,m_i\}; \, i=1,2$, where
$n'_i$ is the momentum quantum number. The inner quantum number $j$ and orbital quantum number  $l$ are
determined by strictly negative (for the eigenstates of the first type) or positive (for the eigenstates of the second type)  integer quantum number $\kappa$  through the relations: $j=|\kappa|-1/2$ and
$l=\mathrm{sign}(\kappa)(\kappa+1/2)-1/2$, respectively. The notation $| vac \rangle $ stands for photonic vacuum state.

Since the case of the bound-bound transition matrix elements has
been thoroughly investigated in \cite{evtrmatel}, here we will be
concentrated mostly on the explicit calculation of contributions
to the Lamb-shift stemming from relativistic transition matrix
elements between bound and unbound eigenstates. The one-electron
relativistic four-component spinor for the bound eigenstates of
hydrogen-like atoms will be used in the form given by Bethe and
Salpeter \cite{BS} and by Rose \cite{Rose} for the unbound
eigenstates. For reader's convenience, we will also follow
notations of these works \cite{BS,Rose}. From here on the natural
units with Gaussian units will be used: $\hb=c=1$, $e^2=\alpha$.

\section{ Transition matrix element contribution to the Lamb
shift}

The starting point of the present study is the conventional expression
for the hydrogenic atom Lamb shift in the second order, resulting from
Eqs.(\ref{DH}-\ref{vp}),

\ba \Delta E_{s_1}=-\mbox{Re}\left[\fr{i\al}{4 \pi^3}\int d^4
k\fr{1}{{k_0}^2-\vk^2+i\veps}\langle  s_1 |e^{i\vk\vr}\alpha^\mu\fr{1}{
E_{s_1}-\hat H_D-k^0+i\eta}\alpha_\mu e^{-i\vk\vr} |s_1 \rangle\right].
\label{p2} \ea

\noi Here $\hat H_D$ is the Dirac Hamiltonian of the hydrogenic
atom, $E_{s_1}$ is an eigenvalue belonging to the bound hydrogenic
eigenstate $| s_1\ran$, $\lan \bar s_1|$ is the corresponding Dirac
conjugate eigenstate and the $\alpha$-matrices are defined as
$\alpha^\mu = \gam^0\gam^\mu$,\,$\alpha^0 = \hat I$ .  Inserting a
complete base of hydrogenic atom eigenstates $\hat I = \sum_{s_2} | s_2
\ran\lan s_2|$ wherever necessary, further simplification of
Eq.(\ref{p2}), better suited for numerical calculations to follow,
is straightforward

\ba
\Delta E_{s_1}=-\mbox{Re}\left[\fr{i\al}{4 \pi^3}\dsum_{s_2}\int d^4 k\fr{1}{{k_0}^2-\vk^2+i\veps}
\fr{\langle s_1
|e^{i\vk\vr}\al^\mu|s_2\ran\lan s_2|
\al_\mu e^{-i\vk\vr}
|s_1 \rangle}{ E_{s_1}-E_{s_2}-k^0+i\eta}\right] \label{p3} \ea

\noi with $E_{s_2}$ being the eigenvalue belonging to the
eigenstate $|s_2\ran$, be it
 bound or unbound as well.

As readily seen from (\ref{p3}), the major problem in direct
evaluation of the Lamb shift, as it stands, is posed by the
necessity to calculate the transition matrix elements $\langle s_1
|e^{i\vk\vr}\al^\mu|s_2\ran$ in the plane wave representation. A
host of various efficient techniques for calculation of the
transition matrix elements of all types possible, i.e. the
elements corresponding to transitions between the bound-bound,
bound-unbound and unbound-unbound hydrogenic atom eigenstates, has
already been developed in our previous papers
\cite{hypersecp,hyperfirstp,evtrmatel,fourtr}. In particular, it
was shown that any transition matrix element between hydrogenic
atom bound eigenstates can be expressed explicitly in terms of
analytic formulae containing hypergeometric functions, which
functions depend only on the absolute value of the wave vector
$\vk$ and the quantum numbers of the two states building up the
matrix element in question. The purpose of the present research is
to show that there is no conceptual difference neither in the way
one could handle contributions from  two distinct types of matrix
elements, the bound-bound and the bound-unbound ones, nor in the structure of the corresponding computational algorithms dedicated to this purpose.

\section{Transition matrix elements }

An elementary building block for any matrix element -
bound-bound, bound-unbound and unbound-unbound - is a Fourier
transform of the kind

\ba  U_{ij}(s_1,s_2)=\int\!\!d\vr e^{i\vec k\vec
r}\phi^*_{s_1,i}(\vr\,) \phi_{s_2,j}(\vr\,), \quad i,j = 1,...,4,
\label{77bis} \ea

\noi  where  $ \phi^*_{s_1,i}(\vr\,)$ and $\phi_{s_2,j}(\vr\,)$ may stand
for the  $i$-th or $j$-th spinor component of relativistic eigenfunctions corresponding to the bound or
unbound eigenstates $|s_1\rangle$ and $|s_2\rangle$ of hydrogenic atom. Any transition matrix
element is a linear combination of four  terms of the kind of (\ref{77bis})
with proper coefficients provided by corresponding matrix $\alpha^\mu, \,
\mu=0,1,2,3$.  Due to expansion

\ba e^{i\vk\vr}= 4\pi\dsum_{l=0}^{\infty}\dsum_{m=-l}^{m=l} i^l
j_l(kr)Y^*_{l,m}(\vr)Y_{l,m}(\vk), \label{newrepr1} \ea

\noi where $j_l(kr)$ is a spherical Bessel function

\ba j_l(z)=\sqrt{\fr{\pi}{2z}}J_{l+1/2}(z)\label{sphericalB} \ea

\noi with $J_\nu(z)$ being a Bessel function of the first kind
defined as

\ba J_{n+1/2}(z)= \sqrt{\fr{1}{2\pi z}}\left\{e^{iz}
\ds_{k=0}^{n}\fr{(i)^{-n+k-1} (n+k)!}{k!(n-k)!(2z)^{k}} +
\right. \nn \\
+ e^{-iz}\left. \ds_{k=0}^{n}\fr{(-i)^{-n+k-1}
(n+k)!}{k!(n-k)!(2z)^{k}}\right\}, \qquad n=0,1,2,...
\label{Besselalt} \ea

\noi the expression (\ref{77bis}) can be rewritten as

\ba  U_{ij}(s_1,s_2)=\nn \\=
C(l_1,m_1,t_1;l_2,m_2,t_2)\cdot 4\pi
\dsum_{l=0}^{\infty}\dsum_{m=-l}^{m=l} i^l Y_{l,m}(\vk)\times\nn\\
\times\dint_0^{2\pi}d\phi\dint_{0}^{\pi}d\theta\sin(\theta)Y^*_{l,m}(\vr)Y^*_{l_1,\tm_1}(\vr)Y_{l_2,\tm_2}(\vr)\times
\nn\\ \times \dint_0^\infty dr r^2 j_l(kr)F^i_1(x_1,l_1,r)F^j_2(x_2,l_2,r). \label{expmatrel} \ea

\noi Here $s_1=\{x_1, l_1, \tilde m_1, t_1\}$, $s_2=\{x_2, l_2,
\tilde m_2, t_2\}$ are the corresponding quantum numbers, and
$\tilde m_1,  \, \tilde m_2$ stand for $m_1\pm 1/2$, \, $m_2\pm
1/2$, where $m_1$,\, $m_2$ are casual magnetic quantum numbers,
$t=1,2$ indicate the type of the state -- the first or the  second
one, $x_i=n'_i$ or $p_i$ for the bound and unbound eigenstates
correspondingly. The coefficient function
$C(l_1,m_1,t_1,l_2,m_2,t_2)$ is a proper product of coefficients
being found at the corresponding spinor components given in
\cite{BS} and \cite{Rose}. The functions $F^i_1(n'_1(or\, p_1)
,l_1,r )$, $F^j_2(n'_2(or\,p_2),l_2,r)$ stand for the
corresponding radial parts in expressions outlined in \cite{BS}
for the bound eigenstates and in \cite{Rose} for the unbound ones.
The integrals over three spherical harmonics in (\ref{expmatrel})
are known as the Gaunt coefficients \cite{Gaunt}. Gaunt
coefficients are closely related to the Clebsch-Gordon
coefficients \cite{Messiah}. They can  be either evaluated
directly by some fast computer algebra algorithm \cite{algGaunt}
or, alternatively, the expression (\ref{expmatrel}) can be
transformed into

  \ba
U_{ij}(s_1,s_2)=C(l_1,m_1,t_1;l_2,m_2,t_2)\cdot 4\pi\times \nn \\
\times \dsum_{l=|\tm_2-\tm_1|}^{l_1+l_2} \fr{i^l}{(2\pi)^{3/2}}
\sqrt{\fr{2l+1}{2} \fr{(l-m)!}{(l+m)!}} \sqrt{\fr{2l_1+1}{2}
\fr{(l_1-\tm_1)!}{(l_1+\tm_1)!}} \sqrt{\fr{2l_2+1}{2}
\fr{(l_2-\tm_2)!}{(l_2+\tm_2)!}}\times\nn\\
\times
Y_{l,m}(\vk)\dint_{0}^{\pi}d\theta\sin(\theta)P_l^m(\cos(\theta))
P_{l_1}^{\tm_1}(\cos(\theta))P_{l_2}^{\tm_2}(\cos(\theta))\Big|_{m=\tm_2-\tm_1}
\times\nn\\
 \!\!\!\times \dint_0^\infty dr r^2 j_l(kr)F^i_1(n'_1(or\,
p_1),l_1,r)F^j_2(n'_2(or\,p_2),l_2,r),\phantom{ ssssss   }
\label{expmatrel1} \ea

\noi written in terms of the Gaunt integrals over three associated
Legendre functions $P_{l}^{m}(x)$:

\ba \dint_{-1}^{1}dxP_{l_1}^{m_1}(x)
P_{l_2}^{m_2}(x)P_{l_3}^{m_3}(x). \label{Gaunt}\ea

\noi Formulas for evaluating the definite integral (\ref{Gaunt})
for $m_1=m_2+m_3$ were derived by Gaunt \cite{Gaunt} and later
confirmed by Infeld and Hull \cite{IH}. In particular, it has been
shown that in order for the integral (\ref{Gaunt}) to be
nonvanishing the triangular conditions

\ba l_1+l_2+l_3 =\mbox{even integer} \label{cond1} \ea

\noi and

\ba l_2-l_1 \le l_3 \le l_2+l_1 \label{cond2} \ea

\noi must be satisfied. However, these conditions are not
exhaustive and the integral (\ref{Gaunt}) may in some cases vanish
even if these conditions are satisfied \cite{addcond}. For the
sake of simplicity and convenience, we would prefer to take
recourse to the standard computer algebra algorithms built in {\it
Mathematica} software package to integrate over the angular
variable $\theta$ analytically, so that the only integration left
in (\ref{expmatrel1}) will be that one over $r$.

\subsection{Radial part integration for bound-bound matrix elements}

If both functions in the integrand, $F^i_1$ and $F^j_2$, belong to the
bound eigenstates and, being such, are expressed each in terms of
Laguerre polynomials after \cite{Davis}, then the resulting
term-by-term evaluation deals with a sum of integrals of the type

\ba \dint_0^\infty dr r^2\sqrt{\fr{\pi}{2kr}}J_{l+\fr{1}{2}}(kr)
e^{-\fr{Zr}{N(n_1',l_1,t_1)a_0}} e^{-\fr{Zr}{N(n_2',l_2,t_2)a_0}}
r^{\gamma_1(l_1)-1+q_1}r^{\gamma_2(l_2)-1+q_2},\ea

\ba \gamma(l)=\sqrt{\kappa(l)-(Z\al)^2},\,\, q_1, q_2 = 0,1,2,
..., \ea

\noi each of which can be expressed explicitly in terms of
hypergeometric functions as

\ba \int_0^\infty dr e^{-\al r}r^{\mu-1} J_{\nu}(\beta
r)=\nn\\
\fr{ \left( \fr{\beta}{2\al}\right)^\nu
\Gam(\nu+\mu)}{\al^\mu\Gam(\nu+1)}\phantom{ }_2F_1\left(
\fr{\nu+\mu}{2},\fr{1+\mu+\nu}{2};\nu+1;-\fr{\beta^2}{\al^2}
\right),\label{142bisbis}\ea

\noi where

\ba [\mbox{Re}(\nu+\mu)>0, \, \mbox{Re}(\al+i\beta)>0, \, \mbox{Re}(\al-i\beta)>0].
\label{143bis}\ea

\subsection{Radial part integration for bound-unbound matrix elements}

In the case when one of the states belongs to the discrete and the
other to the continuous spectrum, a typical integral to be
evaluated is

\ba \dint_0^\infty dr r^2\sqrt{\fr{\pi}{2kr}}J_{l+1/2}(kr)
r^{\gamma(l_1,t_1)-1+q_1} e^{-\fr{Zr}{N(n_1',l_1,t_1)a_0}}\times\nn \\
 \times r^{\gamma(l_2,t_2)-1}
e^{-ipr}\phantom{ }_1F_1(\gam(l_2,t_2)+1+i\al ZW(p)/p,
2\gam(l_2,t_2)+1; 2ipr), \nn \\ q_1=0,1,2,... ; \,\,\,
|\tm_2-\tm_1|\le l\le l_1+l_2,\label{pr1} \ea

\noi where $\phantom{ }_1F_1$ is a confluent hypergeometric
function, $W(p)=\sqrt{m^2+p^2}$. After substituting
expansion(\ref{Besselalt}) for the Bessel function in (\ref{pr1}),
any such integration can be carried out analytically with the
results expressed in terms of hypergeometric functions $\phantom{
}_2F_1$ with the help of two general integration rules

\ba \dint_0^\infty dt e^{-st}t^{b-1}\phantom{}_1F_1(a;c; pt)=
\fr{\Gamma(b)}{s^{b}} \phantom{}_2F_1(a,b;c; p/s),\label{usrel}
\ea

\noi and

\ba \dint_0^\infty dr e^{-sr}r^{b-1}\phantom{}_1F_1(a;c;
pr)\fr{sin(kr)}{kr}=\nn \\
=\fr{\Gam(c)}{\Gam(a)} \sqrt{\fr{2\pi}{k}}\fr{ \left(
\fr{k}{2s}\right)^{1/2} \Gam(1/2+b)}{s^b\Gam(1/2+1)}\phantom{
}_2F_1\left( \fr{1/2+b}{2},\fr{1+b+1/2}{2};1/2+1;-\fr{k^2}{s^2}
\right)+\nn \\
+\fr{1}{2ik}\fr{\Gam(c)}{\Gam(a)}\fr{\Gamma(b-1)}{(s-ik)^{b-1}}\left[
\phantom{ }_2F_1\left(a,b-1;c; \fr{p}{s-ik}\right) -1 \right]- \nn
\\-\fr{1}{2ik}\fr{\Gam(c)}{\Gam(a)}\fr{\Gamma(b-1)}{(s+ik)^{b-1}}\left[\phantom{
}_2F_1\left(a,b-1;c; \fr{p}{s+ik}\right)
   - 1 \right]. \label{usrel2} \ea

 \noi Here, in both cases,

\ba \mbox{Re}[b]>0,\quad \mbox{Re}[s]> 0. \ea

\noi The rules can be readily derived by taking into account the
basic definitions and general properties of the hypergeometric
functions (see, e.g.,\cite{Ryzhik}).


\section{Integration over the wave-vector $\vk$}

The following simplified formulae for the Lamb shift

\ba
\Delta E_{s_1}=-\fr{\al}{4
\pi^2}\dsum_{s_2} {\cal P}\dint_0^\infty dk k\dint d\Omega_{\vk}
 \fr{\langle s_1 |e^{i\vk\vr}\al^\mu|s_2\ran \lan s_2|
\al_\mu e^{-i\vk\vr} |s_1 \rangle }{ E_{s_1}-E_{s_2}-k+i\eta},  \label{p4}
\ea

\noi which results immediately from the original expression
(\ref{p3}) after conventional integration over $k^0$ in the
complex plane, must be employed  in the plain-style numerical
calculation of the contributions from various transition matrix
elements between hydrogenic atom bound eigenstates.  Due to
universal analytic representation for all transition matrix
elements given by expressions of the type (\ref{expmatrel1}), the
integration over the angular polar coordinates $\int
d\Omega_{\vk}$ can be carried  out analytically resulting in
selection rules of the kind

\ba \dint d\Omega_{\vk}
Y^*_{l',m'}(\vk)Y_{l,m}(\vk)=\delta_{ll'}\delta_{mm'}, \ea

\noi
 and only the
remaining one-dimensional principal value integration over $k$ is
to be done numerically. Another performance-enhancing distinctive
feature of the numerical technique introduced in this study is
that, unlike to the numerical approach undertaken in
\cite{Seketm}, a complete matrix element is
actually never calculated at once but rather is broken down into a
sum of very simple easy-to-handle terms of similar structure.
Then, the numerator in Eq.(\ref{p4}) for each intermediate quantum
eigenstate $|s_2\ran$ involved and each $\vk$ would be a sum of four
structurally similar terms

\ba |\langle s_1 |e^{i\vk\vr}|s_2\ran|^2-|\langle s_1
|e^{i\vk\vr}\alpha^1|s_2\ran|^2- |\langle s_1
|e^{i\vk\vr}\alpha^2|s_2\ran|^2-|\langle s_1
|e^{i\vk\vr}\alpha^3|s_2\ran|^2, \label{p5} \ea

\noi each of which is, in its turn, an absolute value  of the sum of four
elementary terms of the kind of (\ref{77bis}). Of course, for
the sake of convenience, the terms inside these sums are broken
down even further by means of the expansion (\ref{Besselalt}) for
the Bessel function and polynomial expansion after Davis
\cite{Davis} for the confluent hypergeometric functions featured
in radial parts of all bound eigenstates.  Afterwards, each thus
built smallest partial term in the numerator is integrated over
$k$ with subsequent summation of all such partial outcomes of this
integration, which makes inherently complicated numerical
calculations feasible even on a modest desktop computer.

\section{Complete renormalization  }

 Our approach to renormalization, developed by J. Seke \cite{Sec1,Sec2}, originates from the concept postulating that the free electron, by its very definition, may not experience any energy shift by the interaction with the vacuum radiation field. This means that all corrections coming from the interaction with the vacuum radiation field for a freely propagating electron have to be removed by renormalization. This should be applied for the renormalization of the bound electron by identifying the free electron contrubitions and removing them.

\subsection{Renormalization of the bound electron to the second order}

The expression for the second order self energy of the bound
electron can be written as \ba
    \overline{\langle  n |} \Sigma^{(2)}_{\rm bnd}(E_n) | n \rangle = ie^2 \overline{\langle n |} \int \frac{d^4 k}{(2\pi)^4k^2} \gamma^\mu \frac{1}{{\Pi\!\!\!/}_n - k\!\!\!/ - m} \gamma_\mu | n \rangle,
\ea where ${\Pi\!\!\!/}_n = \gamma_0 E_n - \gamma_i P^i - e
\gamma_0 A^0$, the on-shell momentum operator of an electron in
the potential $A^0$ and $n$ stands for all quantum numbers
characterizing the corresponding state. In order to identify the
part belonging to the free electron, it seems to be self-evident
to perform a potential expansion and identify the first (zero
potential) term with the contribution of the free electron.
However, it turns out that parts of this term cancel out against
those of all the multipotential terms. Therefore, we perform a
"shifted" potential expansion which automatically cancels all
these terms. This expansion  can be performed by inserting the
complete basis \ba
    \mathbf{1}=\int d^3 p' \sum_{s'} \gamma_0 u_{s'}(\vec p') u^\dag_{s'} (\vec p') |\vec p'\rangle \langle \vec p' | \gamma_0=\int d^3 p' \sum_{s'} |\vec p's'\rangle \overline{ \langle\vec p's' |},
\ea in the expression below \ba
    \overline{\langle  n |} \Sigma^{(2)}_{\rm bnd}(E_n) | n \rangle = ie^2 \overline{\langle  n |} \int \frac{d^4 k d^3 p'}{(2\pi)^4k^2} \sum_{s'}  |\vec p's'\rangle \overline{\langle \vec p's' |} \gamma^\mu \frac{1}{{\Pi\!\!\!/}_n - k\!\!\!/ - m} \gamma_\mu | n \rangle,
\ea accompanied by the shifted potential expansion of the inverse
operator \ba
     \frac{1}{{\Pi\!\!\!/}_n - k\!\!\!/ - m} &=& \frac{1}{{P\!\!\!/}_{\vec p's'} - k\!\!\!/ - m} - \nonumber \\
     &\phantom{=}&- \frac{1}{{P\!\!\!/}_{\vec p's'} - k\!\!\!/ - m} \gamma_0 ( E_n - E_{\vec p's'}- e A^0)\frac{1}{{\Pi\!\!\!/}_n - k\!\!\!/ - m},
\ea which results in \ba
    \overline{\langle  n |} \Sigma^{(2)}_{\rm bnd}(E_n) | n \rangle &=& ie^2 \overline{\langle  n |} \int \frac{d^4 k d^3 p'}{(2\pi)^4k^2} \sum_{s'}  |\vec p's'\rangle \langle\overline{ \vec p's' }| \bigg( \gamma^\mu \frac{1}{{P\!\!\!/}_{\vec p's'} - k\!\!\!/ - m} \gamma_\mu  - \nonumber \\
    &-\gamma^\mu &\!\!\!\frac{1}{{P\!\!\!/}_{\vec p's'} - k\!\!\!/ - m} \gamma_0 ( E_n - E_{\vec p's'}- e A^0)\frac{1}{{\Pi\!\!\!/}_n - k\!\!\!/ - m}\gamma_\mu \bigg) \!| n \rangle.
\ea Here, the first term can immediately be recognized as \ba
    \overline{\langle n |} \int d^3 p' \sum_{s'}  |\vec p's'\rangle \overline{\langle \vec p's' |} \Sigma^{(2)}_{\rm fr}(E_{ \vec p's' }) | n \rangle = \delta m \overline{\langle  n |}n \rangle,
\ea i.e. the so-called mass-renormalization term of the free
electron which, consequently, has to be removed by
renormalization.

The second term, however, has to be treated further by commuting
the expression $ E_n - E_{\vec p's'}- e A^0$ to the left, giving
rise to a commutator of the denominator with the potential $A^0$
and letting it act on the states, where the advantage of the
relation \ba
    \overline{\langle  n |} ( E_n - E_{\vec p's'}- e A^0) |\vec p's'\rangle = \langle n | \gamma_0 ( E_n - E_{\vec p's'}- e A^0) |\vec p's'\rangle,
\ea \ba
    -\gamma_0 E_{\vec p's'} |\vec p's'\rangle = - (\gamma_i p'^i + m) |\vec p's'\rangle = - (\gamma_i P^i + m) |\vec p's'\rangle
\ea is taken. Therefore, \ba
    \overline{\langle  n |} ( E_n - E_{\vec p's'}- e A^0) |\vec p's'\rangle = \langle n |  ( \Pi\!\!\!/_n - m) |\vec p's'\rangle=0,
\ea which means that only the commutator involving the potential
$A^0$ remains, and, being so, cannot be attributed to the free
electron.

Hence, we may conclude that there are no other free-electron
contributions beside the free-electron mass-renormalization term
($\delta m$) in the case of the second-order self-energy. In other
words, this is the only term that has to be removed by the
renormalization, in both cases - the complete and the conventional
renormalization. As a consequence,  the renormalized second-order
bound-electron self-energy reads as \ba
    \overline{\langle  n |} \Sigma^{(2), {\rm ren}}_{\rm bnd}(E_n) | n \rangle &=& \overline{\langle  n |} \big( \Sigma^{(2)}_{\rm bnd}(E_n)  - \delta m \big) |n \rangle = \nonumber \\
    &=&  ie^2 \overline{\langle  n |} \int \frac{d^4 k}{(2\pi)^4k^2} \bigg( \gamma^\mu \frac{1}{{\Pi\!\!\!/}_n - k\!\!\!/ - m} \gamma_\mu -\nonumber \\
     &\phantom{=}&- \int d^3 p' \sum_{s'} |\vec p's'\rangle \langle\overline{ \vec p's' }| \gamma^\mu \frac{1}{{P\!\!\!/}_{\vec p's'} - k\!\!\!/ - m} \gamma_\mu\bigg) | n \rangle.\label{regul}
\ea

\subsection{Numerical renormalization}

 As a result of the presented matrix element calculation analytic technique, only one improper integration over $|{\bf k}|$ is to be carried out numerically with an infinite upper limit, when calculating
contributions of the bound-bound and bound-unbound matrix elements
to the unrenormalized Lamb shift at the second order of the conventional perturbational QED approach.
It is worth noticing, that all the techniques developed above can be applied unchanged to the case when all calculations are carried out with the regularized photon propagator

\ba D_{\mu\nu}^{reg} = -g_{\mu\nu}\left[
\fr{1}{{k_0}^2-\vk^2+i\veps}
-\fr{1}{{k_0}^2-\vk^2+i\veps-\Lambda^2} \right]  \label{regul1}
\ea \noi with $\Lambda$ being the regularization parameter. In
this case the renormalized shift is

\ba
\Delta E^{ren,reg}_{s_1}=\Delta E^{reg}_{s_1}-\delta m\bar{\langle s_1} |s_1 \rangle ,  \label{renorm}
\ea
\noi
where $\delta m$ can be defined through one-dimensional integration  of the kind

\ba
\delta m = \dint_0^\infty dk f(k,\Lambda^2), \label{dm}
\ea

\noi convergent for finite $\Lambda^2$, where $f(k,\Lambda^2)$ is a well-known, albeit cumbersome, function (see, e.g., \cite{Schweber}).
\noi
   Therefore, the renormalization procedure can be accomplished numerically by including the subtraction of the integrand in  Eq.(\ref{dm}) directly in the numerical one-dimensional integration over $|{\bf k}|$ in the regularized expression (\ref{p4}) for the unrenormalized Lamb shift.

\section{Summary}

Our objective so far has been and is to carry out the plain-style
calculations of the second-order renormalized expression
(\ref{p2})  without making any approximations concerning the
transition matrix elements. The next goal would be to carry out
higher-order Lamb shift calculations  by using the shifted
potential expansion and complete renormalization method - both
developed by Seke. It was demonstrated  that the technique of the
transition-matrix element calculation developed so far
\cite{hypersecp,hyperfirstp,evtrmatel,fourtr} and  rectified
further, especially in the present work, fits these purposes quite
well. It is possible now to represent all matrix elements in a
form suitable for extensive numerical calculations with arbitrary
precision. In our approach, all bound-bound as well as
bound-unbound matrix elements are derived analytically. The same
technique of the matrix-element evaluation is applicable for all
possible values of discrete and continuous quantum numbers of
hydrogenic eigenstates and photon wave-vector $\vk$. Only one
improper integration over $|{\bf k}|$ is to be carried out
numerically with an infinite upper limit, when calculating
contributions of the bound-bound and bound-unbound matrix elements
to  the Lamb shift. Matrix elements of both types are represented
now in a closed analytic form suitable for subsequent numerical
calculations of the renormalized Lamb shift as well as for other
possible applications in QED and related fields of research
whatever they might be. Numerical computations, based on our
analytic matrix-element
 representations, can rely totally on functions already
built into the standard computational packages like {\em
Mathematica}, which feature renders this approach  highly
favorable for programming of computationally efficient and
numerically precise algorithms.

\section*{Acknowledgements}

This work was supported by the {\em Dr. Anton Oelzelt-Newinsche
Stiftung} of the Austrian Academy of Sciences. A.V.S. acknowledges
support from the Ausseninstitut of the Technische Universit\"at
Wien, the RFBR grant \mbox{No.09-01-00086-a}  and the RAS research
program "Mathematical Methods in Nonlinear Dynamics".

\end{document}